\documentclass[aps,twocolumn,amsmath,amssymb,preprintnumbers,floatfix,prb,superscriptaddress,longbibliography]{revtex4-2}

\usepackage{comment}
\usepackage[version=4]{mhchem}
\usepackage[utf8]{inputenc}
\usepackage{newtxtext}
\usepackage[upint]{newtxmath}
\usepackage{microtype}
\usepackage{textcomp}
\usepackage{dsfont}
\usepackage{eucal}
\usepackage{siunitx}
\usepackage{soul}

\usepackage{enumerate}
\usepackage{amsfonts}
\usepackage{color}
\usepackage{soul}

\usepackage{todonotes}
\presetkeys%
    {todonotes}%
    {inline}{}

\usepackage{graphicx}

\usepackage[colorlinks,allcolors=blue]{hyperref}
\usepackage[capitalize]{cleveref} 
\usepackage{cleveref}

\definecolor{DarkBlue}{rgb}{0,0,0.80}
\definecolor{DarkRed}{rgb}{0.80,0,0}
\definecolor{Purple}{rgb}{0.55,0,0.55}
\definecolor{Purple}{rgb}{0,0,0.8}

\newcommand{\eg}{e.g.\ }

\let\epsilon\varepsilon

\begin{document}

\title{Josephson transistor and robust supercurrent enhancement with spin-split superconductors}
\author{Chi Sun}
\affiliation{Center for Quantum Spintronics, Department of Physics, Norwegian \\ University of Science and Technology, NO-7491 Trondheim, Norway}
\author{Jacob Linder}
\affiliation{Center for Quantum Spintronics, Department of Physics, Norwegian \\ University of Science and Technology, NO-7491 Trondheim, Norway}\date{\today}
\begin{abstract}
We theoretically investigate the supercurrent flow in a Josephson junction consisting of two spin-split superconductors combined by a normal metal weak link. The normal metal may be driven out of equilibrium, thus modifying the electron and hole occupation and consequently the supercurrent through the system. Considering first an equilibrium normal metal, we find that increasing the spin-splitting field can enhance the supercurrent strongly for long junctions at low temperatures. In contrast to previous work, this is a much larger enhancement (over 100\%) and it is achieved for both parallel and antiparallel spin-splitting field configuations, making the effect robust. On the other hand, when a gate voltage is applied to drive the system out of equilibrium, we demonstrate a more efficient $\pi$-transition of the supercurrent in terms of a lower transition voltage by tuning the spin-splitting. Moreover, we find the application of temperature bias strongly suppresses the supercurrent, resulting in very sharp supercurrent jumps as outputs. 
\end{abstract}

\maketitle
\section{Introduction}

Spin-split superconductors \cite{meservey_prl_70, meservey_physrep_94} have in recent years been shown to yield many interesting phenomena \cite{bergeret_rmp_18}, including giant thermoelectric effects \cite{Ozaeta2014Feb,Bathen2017Jan}, non-local conductances with spatially separated and entangled electrons and holes \cite{hubler_prl_12}, long spin accumulation lengths \cite{hubler_prl_12,quay_nphys_13}, spin valve effects \cite{DeSimoni2018Oct}, efficient spin-charge conversion \cite{jeon_acsnano_21, kamra_prl_24}. The spin-splitting can be achieved by either applying a strong in-plane magnetic field \cite{hubler_prl_12,quay_nphys_13} or by attaching the superconductor to a magnetic material. Due to the spin-splitting, the density of states in the superconductor acquires a large spin-dependent particle-hole asymmetry, which can be leveraged for potential applications in superconducting spintronics and thermoelectronics.

The Josephson junction \cite{josephson_pl_62} forming a weak link between two superconductors is a key element of superconducting circuits. One use which is particularly relevant for the present work is that it has been shown to yield transistor-like effects when driven out-of-equilibrium \cite{wilhelm_prl_98, Baselmans1999Jan,Sun2024Oct}. One key result was demonstrated using superconductor/normal metal/superconductor (SNS) transistor, in which supercurrent suppression as well as a sign reversal ($\pi$-transition) was demonstrated by controlling the energy distribution of the current-carrying states in the normal metal through a gate voltage \cite{Baselmans1999Jan,Baselmans2001Jan,Baselmans2002Oct}. In addition, supercurrent enhancements due to quasiparticle cooling, high-voltage $\pi$-state and low-voltage $\pi$-state have been proposed by tuning the control voltage in the high-temperature regime \cite{Giazotto2004Sep}. Various mechanisms have also been demonstrated to modulate the supercurrent in voltage-controlled Josephson junctions, e.g., the emergence of Majorana bound states \cite{Tiira2017Apr}, trapping of quasiparticles by vortices \cite{Sato2022May,Rocci2021Jan}, electron- and hole-like resonances in the quasiparticle spectrum \cite{Kuhn2001Jan},  quasiparticle injection \cite{bobkova_prb_10, Sankar2022Apr}, and short-range coherent coupling \cite{Matsuo2023Nov}. Besides voltage control, the out-of-equilibrium effects driven by temperature gradient \cite{Fornieri2017May}, spin accumulation \cite{Ouassou2019Sep} and external magnetic flux \cite{Paolucci2023Jan} have also been investigated. 

In this work, we theoretically determine whether the combination of a Josephson junction with spin-split superconductors and a non-equilibrium quasiparticle distribution controllable via different signal inputs can yield interesting effects regarding supercurrent control. The system under consideration is a SNS transistor consisting of two spin-split superconductors with parallel (P) or antiparallel (AP) spin-splitting fields. We report two main results.

\begin{figure}[t!]
    \centering
    \includegraphics[width = 0.3\textwidth]{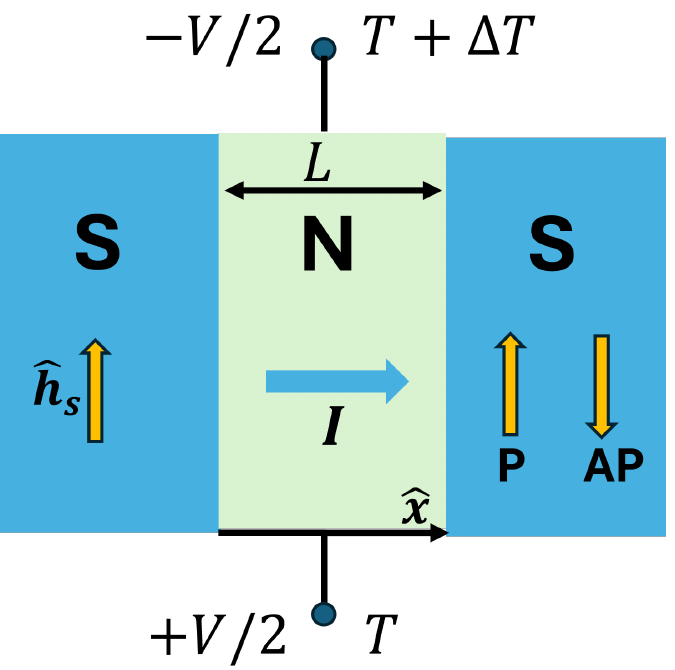}
    \caption{(Color online) SNS Josephson junction considered in this work. The two superconductors are spin-split with the spin-splitting field $\hat{h}_s$, which can be aligned P or AP in the two S layers (as the yellow arrows show). In addition, the transport of charge supercurrent $I$ along $\hat{x}$
can be tuned by a perpendicular voltage or temperature bias.} 
    \label{fig:model}
\end{figure}

Firstly, we consider the system when the electrons in the normal metal weak link are in thermodynamic equilibrium and consider different lengths $L$ of the normal metal and temperatures $T$. We find that, counterintuitively, increasing the spin-splitting field can enhance the supercurrent strongly for long junctions at low temperatures. Such an effect was predicted in a superconductor/insulator/superconductor (SIS) junction in the tunneling limit  \cite{Bergeret2001Apr} for an AP configuration, which we recover in the limit $L \to 0$. In this work, however, we find a much larger enhancement  (over 100\%) when the weak link is a normal metal of length $L/\xi\gg 1$ where $\xi$ is the superconducting coherence length. Interestingly, this takes place both in P and AP configurations, making the effect very robust. 

Secondly, we drive the system out-of-equilibrium by applying three different types of stimuli across the normal metal: electric voltage, temperature difference, and spin accumulation. When the N region is driven out of equilibrium by a voltage, we find that a $\pi$-transition  appears. This sign reversal appears at lower voltages as the junction length or spin-splitting increases. On the other hand, a larger voltage is required to achieve the $\pi$-transition for higher temperatures. We also find that a spin accumulation applied to the normal metal produces the same effect as an electric voltage with respect to the behavior of the supercurrent. In the case of thermal bias, we show that a very small temperature difference applied across a long junction can give rise to sharp supercurrent jumps.

\section{Theory}
In the SNS system (Fig. \ref{fig:model}), quantum transport is described by the Usadel equations \cite{usadel_prl_70} within the typically experimentally relevant diffusive limit. The two spin-split superconductors are treated as reservoirs and the diffusion of the superconducting condensate into the N can be solved by applying appropriate boundary conditions. We then numerically compute the supercurrent across the junction under both equilibrium and nonequilibrium conditions.

\subsection{Usadel equation in the N}
The Usadel equation in N reads
\begin{equation}
    D\nabla(\hat{g}_N^R \nabla \hat{g}_N^R)+i[\epsilon\hat{\rho}_3,\hat{g}_N^R]=0,
\label{eq:Usadel_N}
\end{equation}
in which $D$ is the diffusion coefficient, $\hat{\rho}_3=\text{diag}(1,-1)$ with $1$ being the $2\times2$ unit matrix, and $\epsilon$ represents the quasiparticle energy. Here $\hat{g}_N^R$ denotes the retarded component of the Green function.

Next, we perform the Riccati parametrization \cite{reccati_1995Jul} for $\hat{g}^R_N$:
\begin{equation}
    \hat{g}^R=\begin{pmatrix}
N(1+\gamma\Tilde{\gamma}) & 2N\gamma \\
-2\Tilde{N}\Tilde{\gamma}& -\Tilde{N}(1+\Tilde{\gamma}\gamma)
\end{pmatrix},
\label{eq:reccati_def}
\end{equation}
where the normalization matrices are defined as $N=(1-\gamma\Tilde{\gamma})^{-1}$ and $\Tilde{N}=(1-\Tilde{\gamma}\gamma)^{-1}$. Here the tilde operation corresponds to a combination of complex conjugation $i\rightarrow-i$ and energy reversal $\epsilon\rightarrow-\epsilon$. Compared with the multivalued $\theta$ parameterization \cite{theta_parameterization}, the Riccati parameterization proves highly advantageous for numerical computation due to the parameter bounds for the absolute value of the entries of $\gamma$ 
falling within the range $[0,1]$ for real energies. After applying the Riccati parameterization, Eq. (\ref{eq:Usadel_N}) becomes
\begin{equation}
    D\big(\partial^2_x\gamma_N+2(\partial_x\gamma_N)\Tilde{N}_N\Tilde{\gamma}_N(\partial_x\gamma_N)\big)+2i\epsilon\gamma_N=0,
    \label{eq:Usadel_gamma_N}
\end{equation}
where the transport along the $x$ direction (i.e., perpendicular to the S/N interfaces) is considered. The corresponding equation for $\Tilde{\gamma}_N$ can be
found by taking the tilde conjugate of Eq. (\ref{eq:Usadel_gamma_N}).

\subsection{Usadel equation in the S}
The two spin-split superconducting electrodes are treated as reservoirs. In the presence of spin-splitting, the Usadel equation in S can be given by
\begin{equation}
    [\epsilon\hat{\rho}_3+\hat{\Delta}+\hat{M},\hat{g}_S^R]=0,
    \label{eq:Usadel_S}
\end{equation}
in which $\hat{h}_s$ is the spin-splitting field, and the matrices inside the commutator are
\begin{equation}
    \hat{\Delta}=\begin{pmatrix}
0 & i\Delta\sigma_y\\
i\Delta^{*}\sigma_y& 0 
\end{pmatrix},\quad \hat{M}=\hat{h}_s\cdot \hat{\boldsymbol{\sigma}},\quad {\hat{\sigma}}_n=\begin{pmatrix}
\sigma_n&0\\
0&\sigma_n^{*} 
\end{pmatrix}
\end{equation}
with $\hat{\boldsymbol{\sigma}}=(\hat{\sigma}_x,\hat{\sigma}_y,\hat{\sigma}_z)$ and $\sigma_n$ being the $2\times2$ Pauli matrix. The superconducting gap is described by $\Delta=\Delta(T) e^{i\phi}$, in which the conventional Bardeen-Cooper-Schrieffer (BCS) temperature dependence is utilized, i.e., $\Delta(T)=\Delta_0\tanh({1.74\sqrt{T_c/T-1}})$ with $\Delta_0/T_c\approx1.76$ and $T_c$ is the critical temperature. $\Delta_0$ denotes the superconducting gap amplitude at zero temperature. We have set $k_B=1$. Here we consider $\phi=0$ for the left S while $\phi=\Delta\phi$ for the right S, where $\Delta\phi$ denotes the phase difference between the two superconductors.

Consider $\hat{h}_s=(0,0,h_s)$ corresponding to spin-splitting along the $z$ direction, the analytical solution of Eq. (\ref{eq:Usadel_S}) satisfying the normalization condition $(\hat{g}_S^R)^2=1$ is obtained as \cite{reccati_S_solution}
\begin{equation}
    \hat{g}_S^R=\begin{pmatrix}
c_{\uparrow} & 0 & 0 & s_\uparrow e^{i\phi}\\
0 & c_\downarrow & s_\downarrow e^{i\phi}&0\\
0 & -s_\downarrow e^{-i\phi}&-c_\downarrow&0\\
-s_\uparrow e^{-i\phi}&0&0&-c_\uparrow
\end{pmatrix},
\end{equation}
in which $c_\sigma=\cosh{\theta_\sigma}$ and $s_\sigma=\sinh{\theta_\sigma}$ with $\theta_\sigma=\text{atanh}[\sigma\Delta_0/(\epsilon+\sigma h_s)].$
By equating the above solution with the Riccati parameterization described by Eq. (\ref{eq:reccati_def}), the $2\times2$ matrices $\gamma_S$ and $\Tilde{\gamma}_S$ are solved as
\begin{align}
    \gamma_S=e^{i\phi}\begin{pmatrix}
0 & \frac{s_\uparrow}{1+c_\uparrow}\\
\frac{s_\downarrow}{1+c_\downarrow}& 0
\end{pmatrix},\\
\Tilde{\gamma}_S=e^{-i\phi}\begin{pmatrix}
0 & \frac{s_\downarrow}{c_\downarrow+1}\\
\frac{s_\uparrow}{c_\uparrow+1}& 0
\end{pmatrix},
\end{align}
where $\Tilde{c}_\sigma=c_{-\sigma}$ and $\Tilde{s}_\sigma=s_{-\sigma}$ are employed.

\subsection{Boundary conditions and supercurrent}
Given the expression of $\hat{g}_S^R$, the solution of $\hat{g}_N^R$ can be solved by applying appropriate boundary conditions. At the S/N interface $x=0$, the corresponding Kupriyanov-Lukichev boundary conditions \cite{kupriyanov_zetf_88} are described by
\begin{equation}
    2L_j\zeta_j\hat{g}_j^R\nabla\hat{g}_j^R=[\hat{g}^R_S,\hat{g}^R_N],
    \label{eq:BC}
\end{equation}
in which $j=S(N)$ denotes the superconductor (normal metal) with $\nabla$ being the derivative along the junction $S\rightarrow N$. Here $L_j$ is the
respective length of the materials, and the
interface parameters $\zeta_j = R_B/R_j$ describe the ratio of the
barrier resistance $R_B$ to the bulk resistance $R_j$ of each material. After Riccati parametrization, the
boundary condition at the S/N interface in Eq. (\ref{eq:BC}) becomes
\begin{align}
    \partial_x\gamma_N&=\frac{1}{L_N \zeta_N}(1-\gamma_N\Tilde{\gamma}_S)N_S(\gamma_N-\gamma_S),
\end{align}
and the $\Tilde{\gamma}_N$ counterpart can be found by taking the tilde conjugate of each term. In the following, we use $L$ to denote the N length ($L_N$) since it is the only material length required in our calculation. Note for the N/S interface at $x=L$, $S\leftrightarrow N$ applies for the subscripts in the commutator of Eq. (\ref{eq:BC}).

The supercurrent $I$ flowing in N between the two superconducting
electrodes can in general be expressed as
\begin{align}
   I = \frac{N_0eDA}{8} \int_0^{+\infty} d\epsilon \text{Tr}\{\hat{\rho}_3 (\check{g}_N \partial_x \check{g}_N)^K\},
\end{align}
where $\check{g}_N$ is a $8\times8$ Green function matrix in Keldysh space given by
\begin{equation}
    \check{g}_N=\begin{pmatrix}
\hat{g}_N^R & \hat{g}_N^K\\
0& \hat{g}_N^A
\end{pmatrix}.
\end{equation}
$\hat{g}_N^A$ is the advanced component of the Green function which satisfies $\hat{g}_N^A=-\hat{\rho}_3\hat{g}_N^{R\dagger}\hat{\rho}_3$. $\hat{g}_N^K$ is the Keldysh component of the Green function. $N_0$ is the density of states at the Fermi level
and $A$ is the interfacial contact area. Combing the Drude conductivity described by $\sigma_N = N_0e^2D$ and the normal-state resistance given by $R_N = \frac{L}{\sigma_N A}$, we normalize the $x$-coordinate against $L$ and energy $\epsilon$ against $\Delta_0$ (i.e., $\tilde{\epsilon} = \epsilon/\Delta_0$ and $\tilde{x} = x/L)$ to get the normalized supercurrent expression
\begin{align}
    \frac{eI R_N}{\Delta_0} &= \frac{1}{8} \int_0^{+\infty} d\tilde{\epsilon} \text{Tr}\{\hat{\rho}_3 (\check{g}_N \partial_{\tilde{x}} \check{g}_N)^K\},
\label{eq:normalized_I}
\end{align}
where both sides of the equation are dimensionless.

To compute the supercurrent given the solution of $\hat{g}_N^R$, the general relation for the Keldysh Green function should be used:
\begin{align}
    \hat{g}^K_N = (\hat{g}_N^R \hat{h} - \hat{h}\hat{g}_N^A),
\end{align}
where $\hat{h}$ is the 4$\times$4 distribution function matrix whose explicit expression under different conditions will be given in the following sections. Physically, $\hat{h}$ describes the occupation states of the electrons and holes with various spins in the system, which further provides control over the supercurrent flow across the junction.

\section{Results}

In this section, we numerically compute the supercurrent flow (evaluated arbitrarily at $x=L/2$ since it is conserved throughout the normal metal) 
for both P and AP spin-splitting field alignments in the two spin-split superconductors, by which the device efficiency as a supercurrent transistor will be evaluated. This supercurrent transport along $\hat{x}$ is evaluated in the middle of
the perpendicular control line. Thus, we are analyzing transport in a two-dimensional structure. We first investigate the system under equilibrium conditions and then analyze the non-equilibrium effects for different input signals (voltage, temperature difference, and spin accumulation). We have verified numerically that the supercurrent-phase relation $I(\Delta\phi)$ approaches the generic sinusoidal dependence when the spin-splitting is included for both P and AP configurations (not shown here). This also holds for high voltages and temperature differences \cite{SNS_transistor_1998}, it is therefore assumed that the approximate critical (maximal) supercurrent $I_c=I(\Delta\phi=\pi/2)$ in the following sections. To model inelastic scattering, a small imaginary part $i\delta$ is added to the
quasiparticle energies $\epsilon$ with $\delta/\Delta_0=0.01$.

\begin{figure*}[t!]
    \centering
    \includegraphics[width = 0.97\linewidth]{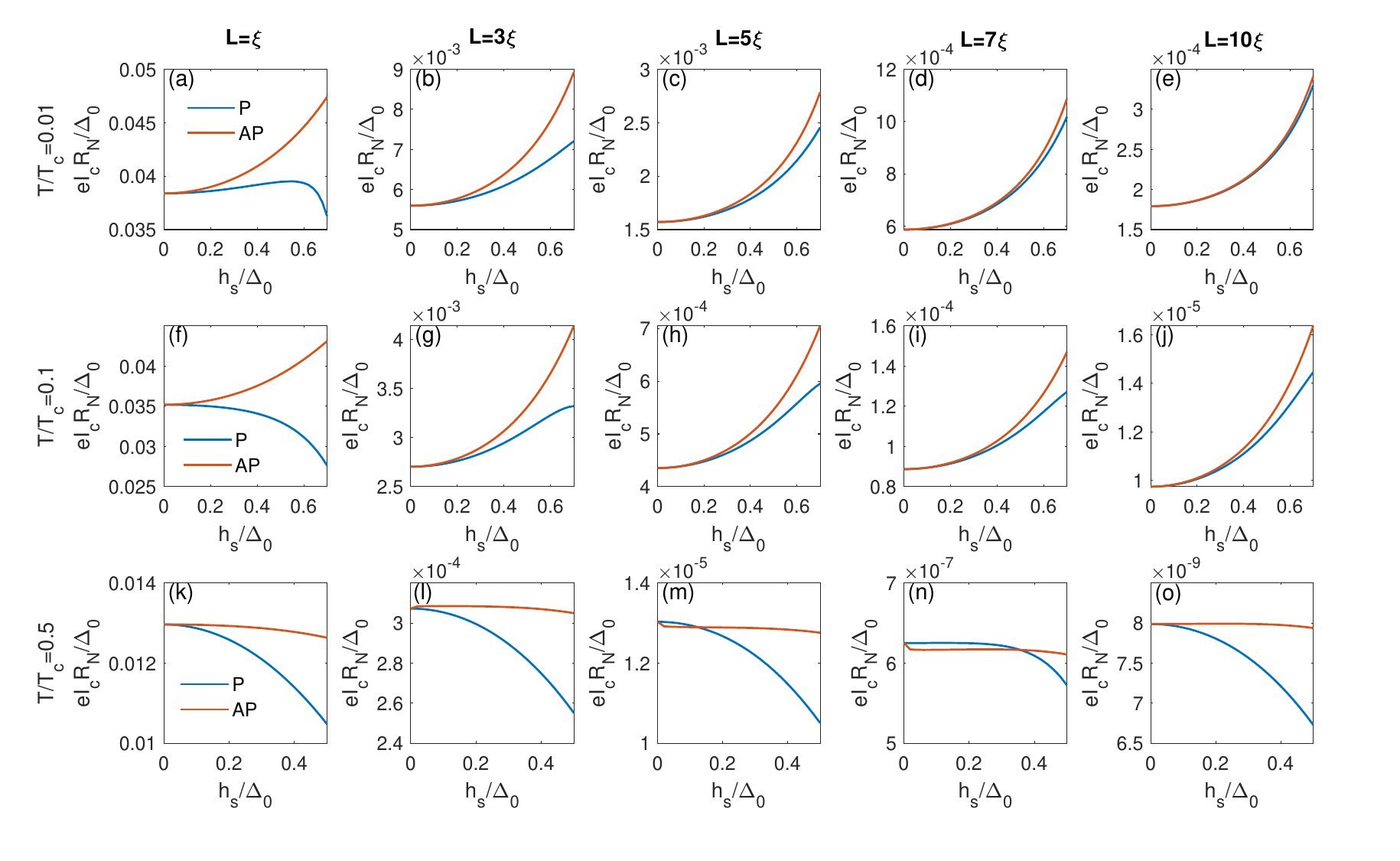}
    \caption{(Color online) The normalized critical supercurrent as a function of the spin-splitting field magnitude $h_s$ for difference temperatures and junction lengths under equilibrium. The conventional BCS temperature dependence is utilized, i.e., $\Delta=\Delta_0\tanh({1.74\sqrt{T_c/T-1}})$ with $\Delta_0/T_c\approx1.76$. Here we use $\zeta_N=5$ and $\delta/\Delta_0=0.01$. } 
    \label{fig:hs_all_new}
\end{figure*}

\subsection{Equilibrium}
For a system in thermal equilibrium, the distribution matrix is given by
\begin{equation}
    \hat{h}=\tanh{(\epsilon/2T)}\hat{\rho}_0,
\end{equation}
in which $\hat{\rho}_0=\text{diag}(1,1)$ with $1$ being the $2\times2$ unit matrix. To study the effect of the spin-splitting, we plot the critical supercurrent $I_c$ as a function of the spin-splitting field magnitude $h_s$ for different temperatures and junction lengths in Fig. \ref{fig:hs_all_new}. Here we stay below the Clogston-Chandrasekhar limit \cite{chandrasekhar_apl_62, clogston_prl_62} $h_s/\Delta_0<0.7$ to avoid the transition to a nonhomogeneous LOFF phase \cite{Fulde1964, Larkin1965}
for larger $h_s$. At temperatures which are not far below $T_c$, a selfconsistent treatment is strictly speaking necessary for the superconducting order parameter $\Delta$, as the phase transition to the normal state occurs at progressively smaller values of $h_s$ as temperature increases (see \eg \cite{ouassou_prb_18}). This only has practical consequence for the plots in the present manuscript where we have set $T/T_c=0.5$, causing us to consider a maximum value of $h_s/\Delta_0 \simeq 0.5$ in that case and thus not including selfconsistency. 

\begin{figure}[b!]
    \centering
    \includegraphics[width = 0.5\textwidth]{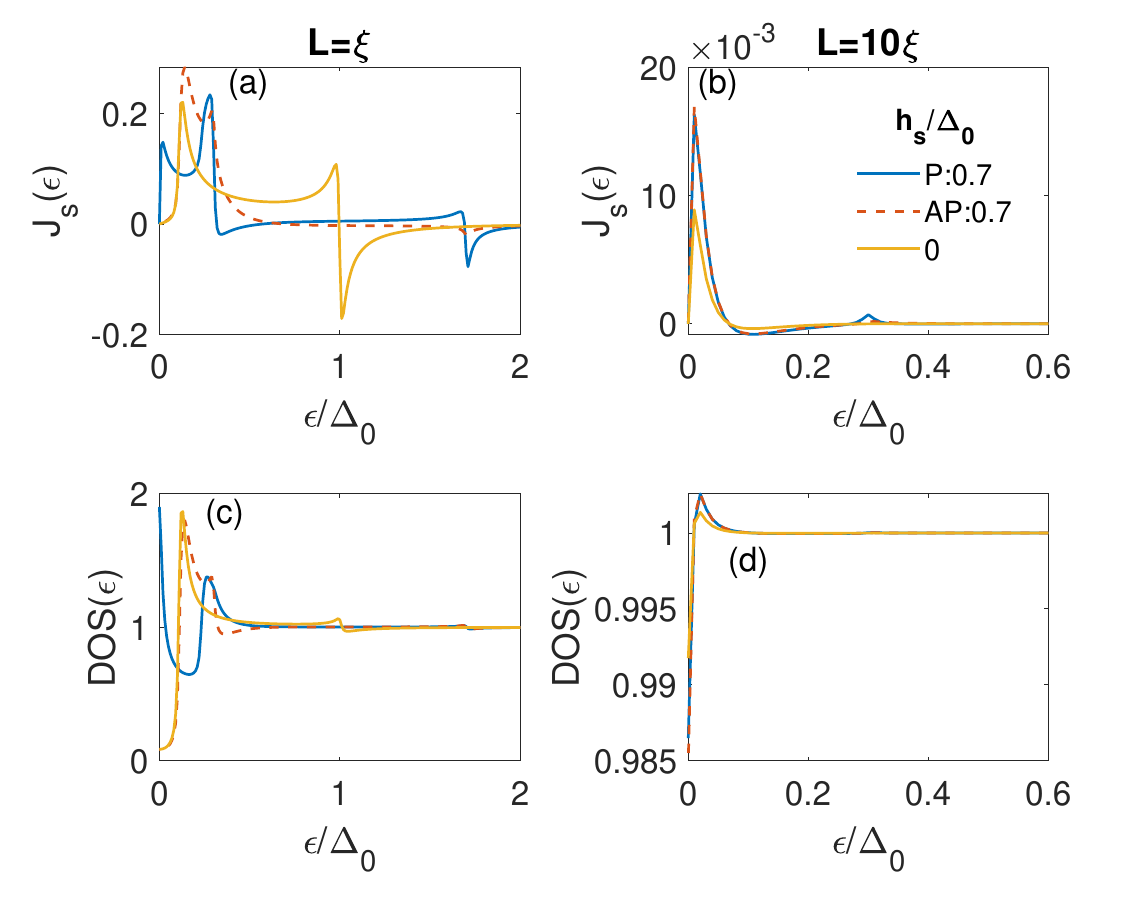}
    \caption{(Color online) The critical spectral current $J_s(\epsilon)$ for (a) $L=\xi$ and (b) $L=10\xi$ at $T/T_c=0.01$ under equilibrium. The corresponding density of states (DOS) for (c) $L=\xi$ and (d) $L=10\xi$.}
    \label{fig:explain_long}
\end{figure}

For a short junction ($L=\xi$) at sufficient low temperature ($T/T_c=0.01$), it is shown in Fig. \ref{fig:hs_all_new}(a) that $I_c$ increases with $h_s$ for an AP alignment while it decreases with $h_s$ for a P alignment. The same phenomenon has been predicted to occur in a S/F(ferromagnet)-I(insulator)-F(ferromagnet)/S Josephson structure \cite{Bergeret2001Apr}, where the supercurrent enhances with the exchange field when the exchange fields (or magnetizations) in the two ferromagnets are antiparallel and vice versa for P orientation at sufficient low temperatures, indicating the analogy between the spin-splitting in our structure and the exchange field in S/F multilayers. As the junction length $L$ increases [Fig. \ref{fig:hs_all_new}(b)-\ref{fig:hs_all_new}(e)], it is found that $I_c$ starts to increase with $h_s$ for P orientation and the corresponding enhancement can approach that for AP alignment when $L=10\xi$. Remarkably, the enhancement reaches a very large value exceeding 100\% at $h_s/\Delta_0\approx0.7$ in Fig. \ref{fig:hs_all_new}(e). This is qualitatively and quantitatively different from Ref. \cite{Bergeret2001Apr}: the supercurrent enhancement occurs at both P and AP configurations and is much larger when considering longer junctions $L/\xi\gg 1$. This can be explained by comparing the corresponding critical spectral currents $J_s(\epsilon)$ [i.e., integrand of Eq. (\ref{eq:normalized_I})] for short and long junctions shown in Fig. \ref{fig:explain_long}. Here, $J_s(\epsilon)$ is strongly modulated by $L$ and is greatly increased at small energies $\epsilon\ll \Delta$ for both P and AP at $L=10\xi$ in the presence of a strong spin-splitting $h_s/\Delta_0=0.7$. Examining the density of states (DOS) for the corresponding system $L=10\xi$ in Fig. \ref{fig:explain_long}(d), we see that the superconducting proximity effect in the normal metal is enhanced at low energies by the spin-splitting field present in the superconductor. The strength of the proximity effect is determined by the deviation of the DOS from its normal-state value of unity. The increase in presence of superconducting correlations gives rise to a larger critical supercurrent sustainable by the system. In addition, it is found that the difference between the magnitude of the supercurrent in the P and AP alignment decreases with increasing junction length $L$, analogously to how the giant magnetoresistance effect decreases with spacer thickness in ferromagnetic spin-valves oriented P or AP \cite{tsymbal_ssp_01}.

The supercurrent enhancement holds when temperature increases slightly [see Fig. \ref{fig:hs_all_new}(f)-\ref{fig:hs_all_new}(j) at $T/T_c=0.1$], showing robust enhancement of supercurrent by involving spin-split superconductors. Note that the situation changes for much higher temperatures [e.g., $T/T_c=0.5$ in Fig. \ref{fig:hs_all_new}(k)-\ref{fig:hs_all_new}(o)], where $I_c$ decreases with $h_s$ for both P and AP orientations regardless of $L$, giving the same trends as shown in Ref. \cite{Bergeret2001Apr} for higher temperatures. Our results show that Josephson junctions with spin-split superconductors even in equilibrium provide new opportunities by considering the long-junction limit.

\subsection{Voltage difference}
Next, we drive the SNS Josephson transistor out-of-equilibrium by applying a voltage difference $V$ (see Fig. \ref{fig:model}), which manipulates the occupation states of electrons and holes in the N with the distribution function \cite{SNS_transistor_1998,Heikkila_distribution}
\begin{equation}
    \hat{h}=\frac{1}{2}\big\{\tanh{[(\epsilon+eV/2)/2T]}+\tanh{[(\epsilon-eV/2)/2T]}\big\}\hat{\rho}_0.
    \label{eq:V_distribution}
\end{equation}
This relation holds near the center of the voltage-biased N and provides a characteristic two-step profile of the occupation of states as a function of energy. We underline that the applied voltage to the N region is not a gate voltage which alters the effective Fermi level, an effect typically utilized in two-dimensional electron gases in semiconducting materials, but instead actually applies a small control current running through the N region transversely to the Josephson junction.

\begin{figure}[t!]
    \centering
    \includegraphics[width = 0.5\textwidth]{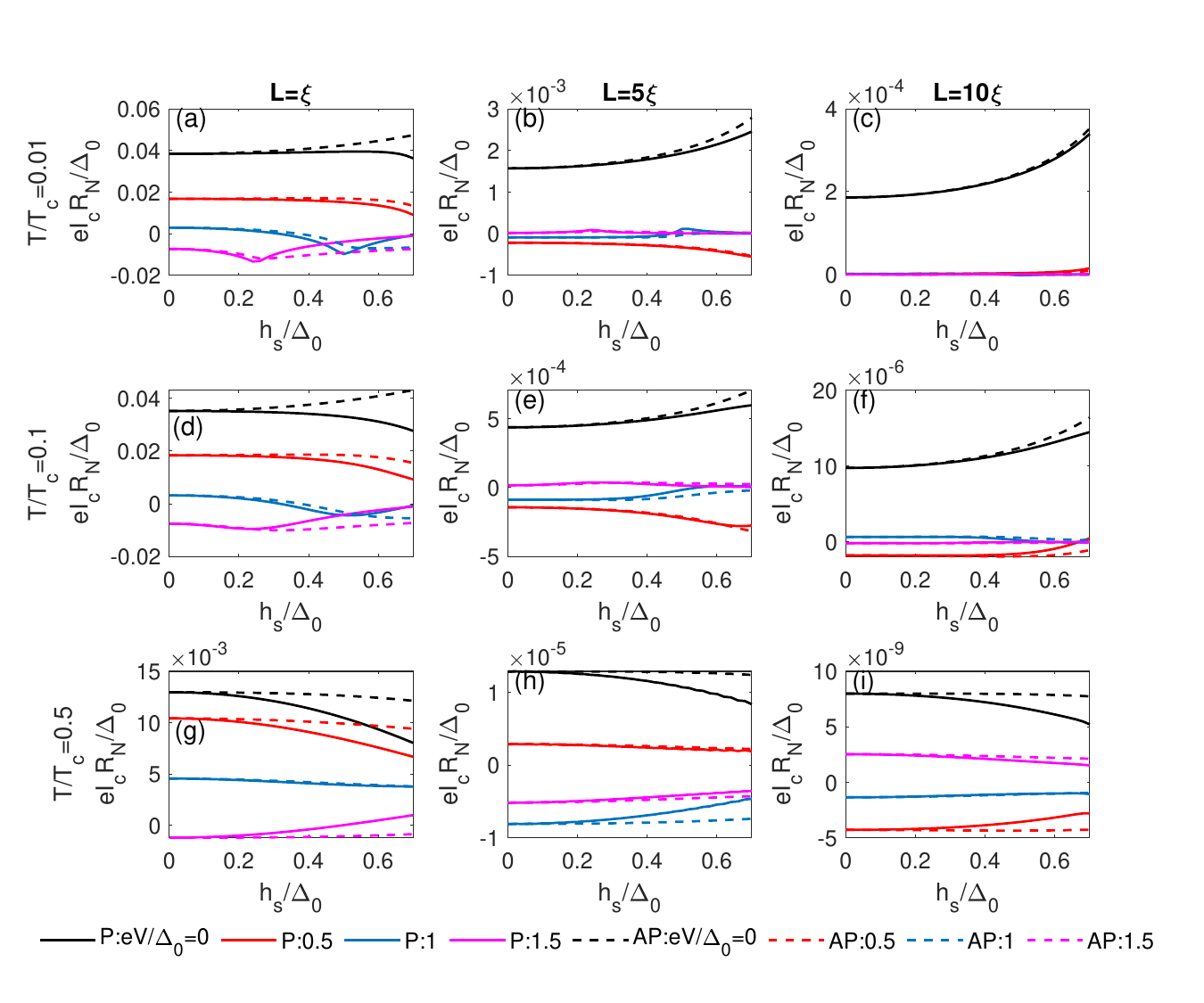}
    \caption{(Color online) The normalized critical supercurrent as a function the spin-splitting field magnitude $h_s$ (both P and AP) for different gate voltages.} 
    \label{fig:hs_V_all}
\end{figure}

\begin{figure}[b!]
    \centering
    \includegraphics[width = 0.5\textwidth]{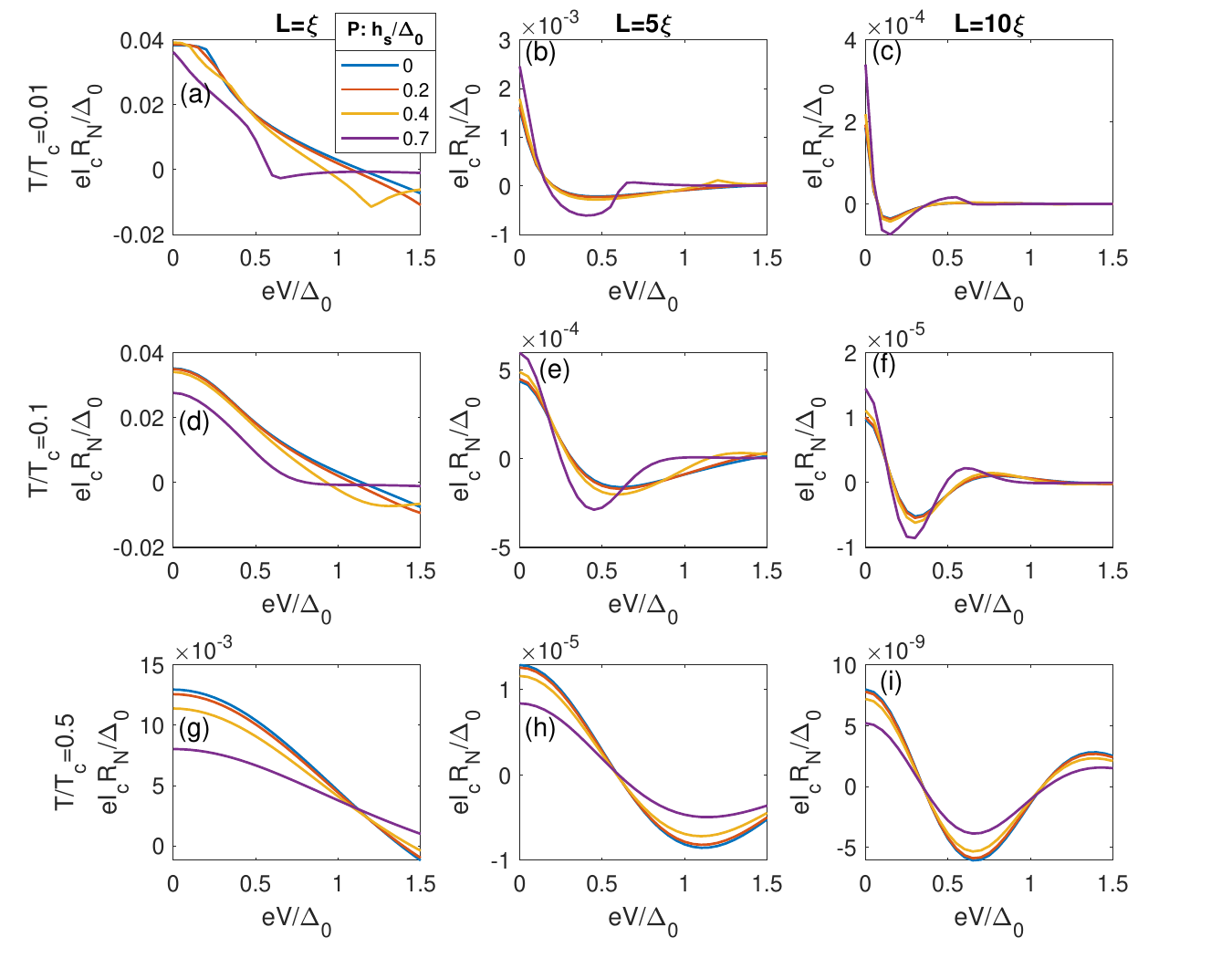}
    \caption{(Color online) The normalized critical supercurrent as a function of voltage for different spin-splitting field magnitudes $h_s$ with P alignment.} 
    \label{fig:Voltage_all_P}
\end{figure}

\begin{figure}[b!]
    \centering
    \includegraphics[width = 0.5\textwidth]{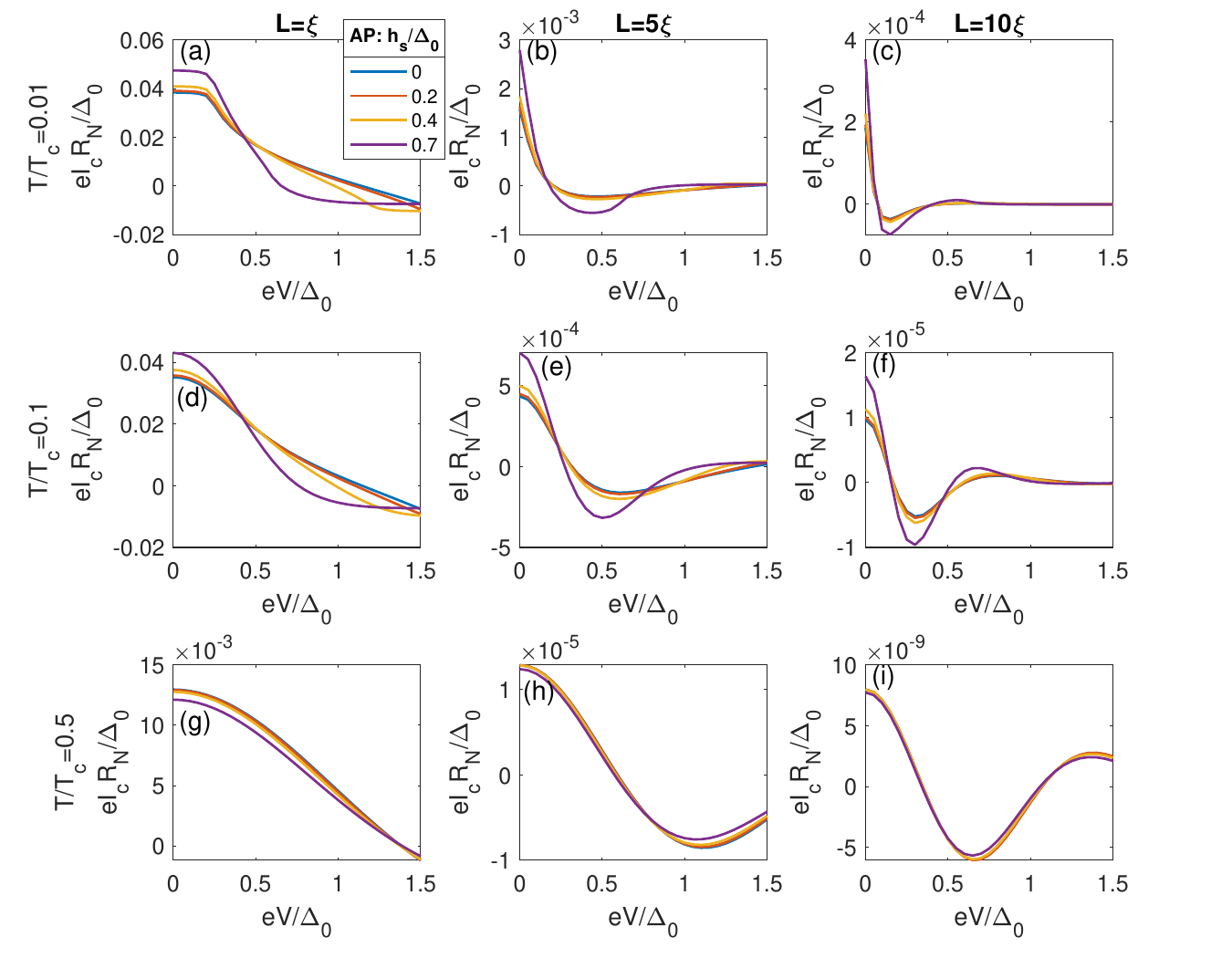}
    \caption{(Color online) The normalized critical supercurrent as a function of voltage for different spin-splitting field magnitudes $h_s$ with AP alignment.} 
    \label{fig:Voltage_all_AP}
\end{figure}

In Fig. \ref{fig:hs_V_all}, we plot $I_c$ as a function of $h_s$ when different gate voltage $V$ values are applied, in which the black curves (solid for P and dashed for AP) for comparison correspond to the equilibrium case without gate voltage as shown in Fig. \ref{fig:hs_all_new}. It can be seen that $I_c$ is generally decreased for both P and AP orientations when $V$ is added, which further gives a sign reversal ($\pi$-transition) of $I_c$ for larger $V$. This supercurrent reduction and $\pi$-transition creates the possibility of using these SNS junctions as a supercurrent transistor \cite{SNS_transistor_1998}. In addition, the introduction of $V$ changes the trend that the supercurrent enhancement with $h_s$ exhibited for longer junctions at low temperatures under equilibrium: the amplitude of $I_c$ can now either decrease or increase with $h_s$ in different regions combined with the sign reversal. To show this clearer, we also plot $I_c$ as a function of $V$ for different $h_s$ in Fig. \ref{fig:Voltage_all_P} and Fig. \ref{fig:Voltage_all_AP} for P and AP alignments, respectively. It reproduces the typical voltage-dependent supercurrent reduction as well as a transition to a $\pi$ junction observed in Refs. \cite{Baselmans1999Jan,Baselmans2001Jan,Baselmans2002Oct} for both P and AP alignments. The supercurrent will finally decay to zero for much larger voltages. It is found that the introduction of $h_s$ decreases the voltage bias required to realize this $\pi$-transition. This is because the existence of $h_s$ splits the original one dip of $I_c$ appearing at higher $V$ without $h_s$ into two dips and the first dip at a lower $V$ appears as the $\pi$-transition here. As for junction length $L$ and temperature $T$ dependences, it is seen that the $\pi$-transition appears at lower (higher) $V$ as $L$ ($T$) increases. Therefore, a long SNS Josephson junction with large spin-splitting at a low temperature is preferable to achieve the typical $\pi$-transition at a very small $V$. Moreover, it is shown in Figs. \ref{fig:Voltage_all_P} and \ref{fig:Voltage_all_AP} that the dips in the voltage-induced $\pi$-transition
become less sharp as the temperature $T$ increases. This can be explained
by the distribution function magnitude in Eq. (\ref{eq:V_distribution}) which varies smoother with energy as $T$
increases.

We have also found that spin accumulation or spin voltage can play exactly the same role as voltage for supercurrent control in our Josephson transistor since they share the same distribution function at the center of the perpendicular control line in N. However, if the measurement is performed at the position deviates from the center, electric voltage and spin accumulation modulate the supercurrent in different ways.

\subsection{Temperature difference}

As an alternative method of driving the itinerant electrons in the N weak link out of equilibrium, we consider an applied temperature difference $\Delta T$ (see Fig. \ref{fig:model}) instead of an electric voltage. The distribution function becomes
\begin{equation}
    \hat{h}=\frac{1}{2}\big\{\tanh{(\epsilon/2T)}+\tanh{[\epsilon/2(T+\Delta T)]}\big\}\hat{\rho}_0,
    \label{eq:T_distribution}
\end{equation}
which is valid near the center of N.

\begin{figure}[t!]
    \centering
    \includegraphics[width = 0.5\textwidth]{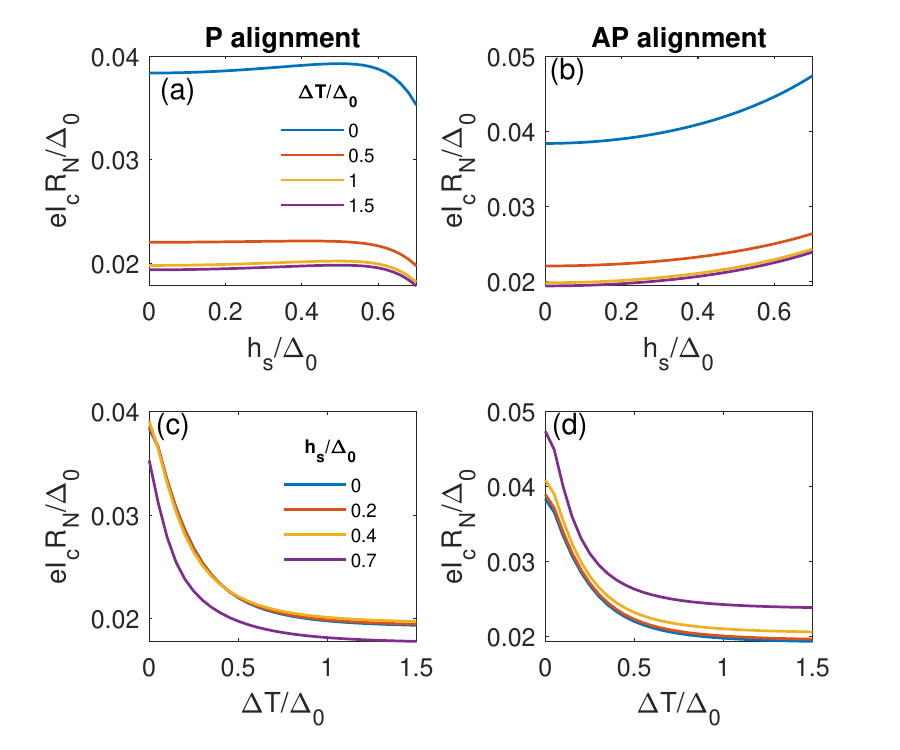}
    \caption{(Color online) Temperature bias dependence of the normalized critical supercurrent at $T/T_c=0.01$ for a short junction ($L=\xi$) with P and AP alignments.} 
    \label{fig:T_all_short}
\end{figure}

In Fig. \ref{fig:T_all_short}, we investigate the temperature difference-induced supercurrent modulation for a short junction ($L=\xi$) at $T/T_c=0.01$. The main effect of $\Delta T$ is to suppress $I_c$ without changing its qualitatibe behavior with respect to $h_s$. In effect, $I_c$ decreases with $h_s$ for P alignment while it increases with $h_s$ for AP alignment, similarly to Fig. \ref{fig:hs_all_new}(a). When plotting $I_c$ as a function of $\Delta T$ for different $h_s$ in Fig. \ref{fig:T_all_short}(c)-\ref{fig:T_all_short}(d), the $\pi$-transition does not appear for $\Delta T/\Delta_0<1.5$, indicating that it is easier to induce the $\pi$-transition via voltage compared to temperature bias. This can be attributed to the larger $\Delta T$-induced distribution magnitude in Eq. (\ref{eq:T_distribution}) at $\epsilon/\Delta_0\rightarrow0$ where the distribution-independent part of the spectral current shows a large positive peak, giving a large positive supercurrent contribution and therefore hindering the appearance of $\pi$-transition.

\begin{figure}[t!]
    \centering
    \includegraphics[width = 0.5\textwidth]{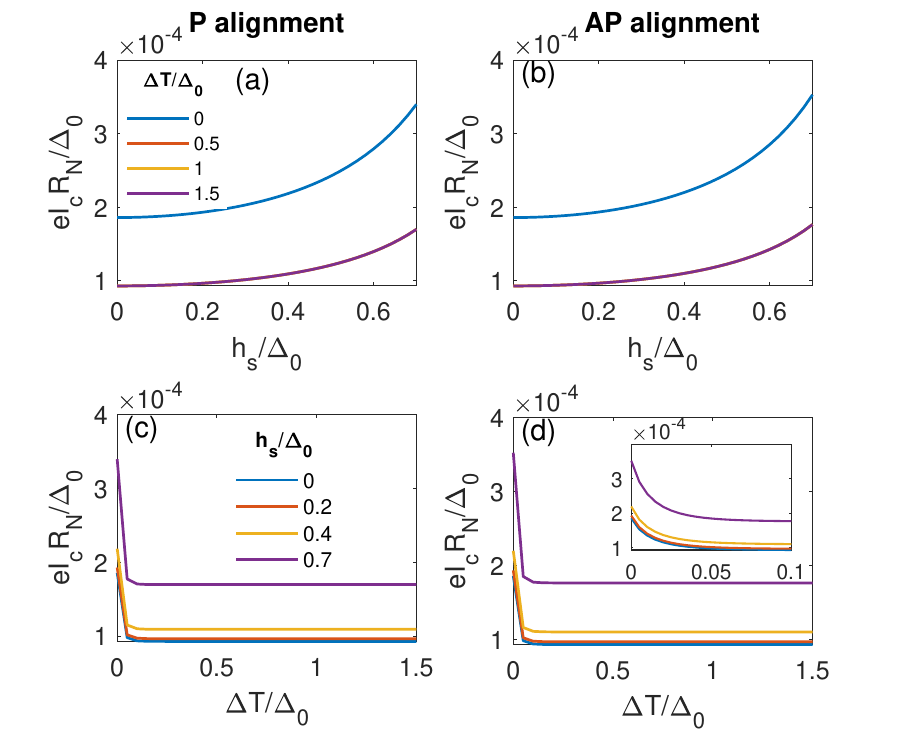}
    \caption{(Color online) Temperature bias dependence of the normalized critical supercurrent at $T/T_c=0.01$ for a long junction ($L=10\xi$) with P and AP alignments. The inset is the enlarged version of its corresponding main panel. Note that the red and yellow curves
almost overlap with the purple one in (a) and (b).}
    \label{fig:T_all_long}
\end{figure}

As for a long junctions ($L=10\xi$) at $T/T_c=0.01$ in Fig. \ref{fig:T_all_long}, we find that the supercurrent enhancement for larger $h_s$, in both P and AP orientations, is maintained when $\Delta T$ is applied. On the other hand, the introduction of $\Delta T$ strongly suppresses $I_c$, giving very sharp jumps in Fig. \ref{fig:T_all_long}(c)-(d) for both P and AP alignments. As shown in the inset of Fig. \ref{fig:T_all_long}(d) for AP alignment, the output signal drops around 50\% already at $\Delta T/\Delta_0 = 0.05$, which corresponds to $\simeq$ 0.05 K for a gap $\Delta_0 = 0.1$ meV which is comparable to the gap in Al. Similar sensitive behavior also holds for the P alignment in Fig. \ref{fig:T_all_long}(c).

\section{Conclusion}
We comment on the possible experimental realizations regarding the parameter values used in our model. The interface parameter $\zeta_N=5$ can correspond
to a interface with low to intermediate transparency that is achievable
experimentally. On the other hand, the relatively small spin-splitting $h_s/\Delta_0<0.7$ is suitable to
be achieved by attaching a magnetic material to the superconductor. Well-defined spin-splitting in a superconductor has been demonstrated
in experiments involving aluminum/europium chalcogenide heterostructures \cite{Strambini2017Oct,Xiong2011Jun}.

In conclusion, we have computed the supercurrent flow in a SNS Josephson junction made of spin-split superconductors connected to a normal metal which can be driven out of equilibrium. Considering the system in equilibrium, we find that a robust supercurrent enhancement with respect to the spin-splitting field can be achieved for longer junctions at low temperatures, regardless of whether the spin-splitting fields in the superconductors are aligned P or AP. The enhancement ratio can exceed 100\% by tuning the spin-splitting. On the other hand, when a gate voltage is applied to drive the system out-of-equilibrium, we show that the typical $\pi$-transition of supercurrent can be achieved and the introduction of spin-splitting further reduces the voltage required for the transition. Moreover, we find the application of temperature bias strongly suppresses the supercurrent, resulting in a very sharp supercurrent jump as a response to small temperature differences.

\begin{acknowledgments}
This work was supported by the Research Council of Norway through Grant No. 323766 and its Centres of Excellence funding scheme Grant No. 262633 “QuSpin.” Support from Sigma2 - the National Infrastructure for High-Performance Computing and Data Storage in Norway, project NN9577K, is acknowledged.
\end{acknowledgments}

\bibliography{bib}

\end{document}